\documentclass[twocolumn]{autart}    % Enable this line and disable the 
\usepackage{graphicx}          % Include this line if your 
\usepackage{amsmath} % assumes amsmath package installed
\usepackage{amsfonts}
\usepackage{graphicx}
\usepackage{amsmath}
\usepackage{color} % PER COMMENTI
\usepackage{booktabs} % DEVO CONTROLLARE SE È IEEE-COMPLIANT
\usepackage{comment}
\usepackage{epstopdf}
\usepackage{caption}
\usepackage{subcaption}
\usepackage{cite}
\usepackage{cases}
\usepackage{dsfont}
\usepackage{empheq}
\newtheorem{assumption}{Assumption}

\newtheorem{theorem}{Theorem}
\newtheorem{remark}{Remark}

\newcommand*{\reply}[1]{{{\color{black} #1}}}

\newenvironment{proof}{\paragraph*{Proof.}}{\hfill$\square$}

\begin{document}
\newcommand{\agent}[1]{\mathbf{p}_{#1}(t)}                                % agent
\newcommand{\agents}{\mathbf{p}(t)}                                % agents
\newcommand{\vagent}[1]{\dot{\mathbf{p}}_{#1}(t)}                                % agent
\newcommand{\N}{N}                      % number of agents
\newcommand{\density}{\rho(\mathbf{x},t)}                % target distribution
\newcommand{\densityd}{\bar \rho(\mathbf{x},t)}                % target distribution
\newcommand{\de}{\mathrm{d}}     % upright d as in dx/dt

\newcommand{\W}[3]{\mathcal{W}_{#1}^{#1}\left(#2, #3\right)}
\newcommand{\WH}[1]{\W{#1}{\mu_t}{\bar \mu_t}}

\newcommand{\dir}[1]{\delta(\mathbf{x}-#1)}
\newcommand{\domain}{\Omega}
\newcommand{\absv}[1]{\left\lvert #1 \right\rvert}
\newcommand{\normp}[2]{\left\lVert #1 \right\rVert_#2}
\newcommand{\lip}[1]{\mathrm{L}_{#1}}
\newcommand{\Rn}{\mathbb{R}}
\newcommand{\indicator}[2]{\mathbf{1}_{#1}\left(#2\right)}
\newcommand{\Lp}[1]{\mathcal{L}^{#1}}
\newcommand{\Lpnorm}[2]{\normp{#1}{\Lp{#2}}}
\newcommand{\Laguerre}[1]{\mathcal{V}_{#1}}
\newcommand{\Voronoi}[1]{\Tilde{\mathcal{V}}_{#1}}

\begin{frontmatter}
\title{Semi-discrete Optimal Transport \\ for Time-Varying Multi-Agent Coverage Control}

\thanks[footnoteinfo]{This paper was not presented at any IFAC 
meeting.\\ $^*$ Corresponding author}

\author[IN]{Italo Napolitano}\ead{i.napolitano@ssmeridionale.it},    % Add the 
\author[MdB]{Mario di Bernardo}\ead{mario.dibernardo@unina.it}               % e-mail address 

\address[IN]{Modeling and Engineering Risk and Complexity, Scuola Superiore Meridionale, Naples, Italy}  % Please supply                                              
\address[MdB]{Department of Electrical Engineering and Information Technology, University of Naples Federico II\\ and Modeling and Engineering Risk and Complexity, Scuola Superiore Meridionale, Naples, Italy}   
% full addresses

%\begin{keyword}                           % Five to ten keywords,  
%Optimal Transport; Coverage Control; Autonomous Systems; Time-Varying Density.
%\end{keyword}                             % keyword list or with the 
                                          % help of the Automatica 
                                          % keyword wizard

\begin{abstract}                          % Abstract of \begin{abstract}
Coverage control algorithms have traditionally focused on static target densities, where agents optimally cover a fixed spatial distribution. However, many applications, such as environmental monitoring, surveillance, and adaptive sensing, involve time-varying densities. While time-varying coverage has been studied in Voronoi-based frameworks, extending recent optimal transport formulations of static coverage control to time-varying target densities remains an open problem.
This paper presents a semi-discrete optimal transport framework for time-varying coverage control, in which agents \reply{track the first-order optimality conditions associated with} minimizing the instantaneous Wasserstein distance from an evolving target density. The proposed approach is based on a coupled system of differential equations governing agent positions and the dual variables defining Laguerre regions. \reply{The resulting optimality residuals converge exponentially to zero, with global convergence established for one-dimensional domains. We also derive decentralized approximations and} numerical simulations demonstrate improved tracking performance over quasi-static and Voronoi-based methods.
\end{abstract}

\end{frontmatter}

\section{Introduction}
The coverage control problem addresses the coordination of mobile agents to achieve an optimal spatial distribution over a domain, by designing control laws that minimize a cost function quantifying how well a finite set of agents covers a continuous spatial distribution. Applications include environmental monitoring, search and rescue, surveillance of moving crowds, adaptive sensor deployment, and probabilistic formation control~\cite{lin2025heterogeneous,bandyopadhyay2014probabilistic}. This problem has received extensive attention over the past two decades~\cite{krishnan2022multiscale,cortes2004coverage,lin2025heterogeneous}, with the seminal work of Cort\'es et al.~\cite{cortes2004coverage} introducing a framework based on Centroidal Voronoi Tessellations (CVTs), in which agents move toward the centroids of their Voronoi regions to minimize locational costs for stationary target densities. 
Many applications require agents to track time-varying densities, including environmental monitoring of dynamic phenomena, surveillance of moving crowds, adaptive sensor deployment, and tracking probability distributions in stochastic systems \cite{lin2025heterogeneous}.
Extensions to time-varying densities have been developed by Lee et al.~\cite{lee2015multirobot} using control laws with approximated temporal derivatives, and by Lin et al.~\cite{lin2025heterogeneous} via time-varying Control Lyapunov Functions, both within the Voronoi-based framework.

Optimal transport (OT) provides an alternative perspective by offering a principled means of comparing continuous and discrete measures through the Wasserstein distance~\cite{villani2008optimal}, i.e., the semi-discrete formulation, making it well suited for coverage control~\cite{krishnan2022multiscale,krishnan2024distributed,inoue2020optimal}. Inoue et al.~\cite{inoue2020optimal} cast static coverage as a semi-discrete OT problem using Laguerre diagrams and the Kantorovich dual formulation, establishing connections between CVT methods and OT, while Krishnan and Martinez~\cite{krishnan2024distributed} developed distributed algorithms with convergence guarantees in the large-scale limit. All these works restrict attention to stationary densities.

This paper develops a semi-discrete OT framework for time-varying coverage control. The task is formulated as a continuous-time, semi-discrete OT problem in which the agents minimize, at each instant, the Wasserstein distance between their empirical measure and a time-varying target measure. Unlike dynamic optimal transport formulations seen in mean-field control~\cite{chen2021optimal}, we consider a limited number of agents and pose the problem of minimizing the \emph{instantaneous} semi-discrete Wasserstein distance, in a formulation that generalizes well-established Voronoi-based approaches and yields a more accurate approximation of the target density than CVT extensions~\cite{lee2015multirobot}. Unlike existing OT approaches~\cite{inoue2020optimal}, which assume stationary target references, we account for a target measure continuously evolving in time, so that the problem fundamentally changes from regulation to tracking. To the best of the authors' knowledge, this is the first optimal transport solution to time-varying coverage control. 
The main contributions are: (i) a formulation of time-varying coverage as a time-dependent semi-discrete optimal transport problem via instantaneous Wasserstein minimization; (ii) coupled dynamics for agents and dual variables, combining proportional feedback with feedforward compensation to achieve exponential tracking of the time-varying Laguerre barycenters; and (iii) \reply{decentralized approximations.} 
Numerical simulations demonstrate superior tracking performance compared to both quasi-static OT and Voronoi-based methods.

Sec.~\ref{sec:prelim} reviews semi-discrete optimal transport and its connection to coverage control. Sec.~\ref{sec:problem} formulates the time-varying coverage problem. Sec.~\ref{sec:solution} derives the proposed control laws. 
Sec.~\ref{sec:validation} provides numerical validation, and Sec.~\ref{sec:conclusions} concludes the paper.

\section{Background}
\label{sec:prelim}
Scalars are lowercase ($x$), vectors bold lowercase ($\mathbf{x}$), and matrices bold uppercase ($\mathbf{A}$). The operator $\text{vec}(\cdot)$ stacks vectors, e.g., $\mathbf{p} := \text{vec}(\mathbf{p}_1, \ldots, \mathbf{p}_N)$. The $N$-dimensional identity matrix is denoted by $\mathbf{I}_N$. Matrices of zeros and ones are denoted by $\mathbf{0}_{N \times M}$ and $\mathbf{1}_{N \times M}$, respectively. For $M=1$, they are $\mathbf{0}_N$ and $\mathbf{1}_N$. Partial time derivatives are $\partial_t$. Target quantities carry a bar, e.g., $\densityd$. The set of Borel probability measures on $\domain$ with finite second moment is $\mathcal{P}_2(\domain)$, and $\mathcal{W}_p(\cdot,\cdot)$ denotes the $p$-Wasserstein distance. \reply{As in \cite{inoue2020optimal}, $\mathcal{W}_2^2$ is the transport value associated with the half-squared Euclidean cost $\tfrac{1}{2}\lVert\cdot\rVert_2^2$.}

\subsection{Semi-discrete Optimal Transport}
\label{sec:sdot}
Optimal Transport compares probability measures by minimizing the cost of redistributing mass between them. The Kantorovich formulation allows probabilistic transport plans, and its optimum induces the Wasserstein distance, $\mathcal{W}_p$~\cite{villani2008optimal}. In the semi-discrete setting, a discrete measure is compared against a continuous one, and under mild assumptions strong duality holds. In this setting, the OT map induces a partition of the domain into Laguerre cells that generalize Voronoi cells through additive weights~\cite{aurenhammer1987power}.

Consider a discrete probability measure $\mu := \frac{1}{N}\sum_{i=1}^N \delta_{\mathbf{p}_i}$ with $\mathbf{p}_i\in \Rn^d$, and an absolutely continuous measure $\bar\mu \in \mathcal{P}_2(\domain)$ with density $\bar\rho(\mathbf{x})$. For the half-squared Euclidean cost, the Kantorovich dual formulation, exploiting the $c$-transform, yields~\cite{villani2008optimal,peyre2019computational}
\begin{equation*}
\mathcal{W}_2^2(\mu,\bar\mu) = \max_{\boldsymbol{\phi}\in\mathbb{R}^N} F(\mathbf{p},\boldsymbol{\phi}),
\end{equation*}
where $\boldsymbol{\phi}:=\mathrm{vec}(\phi_1,\dots,\phi_N)$ is the dual variable and
\begin{multline}\label{eq:dual_objective}
F(\mathbf{p},\boldsymbol{\phi}) := \sum_{i=1}^N \Bigg[
\int_{\Laguerre{i}(\mathbf{p},\boldsymbol{\phi})}
\frac{1}{2}\|\mathbf{p}_i-\mathbf{x}\|_2^2\,\bar\rho(\mathbf{x})\,\mathrm{d}\mathbf{x} \\
+ \left(\frac{1}{N} - \int_{\Laguerre{i}(\mathbf{p},\boldsymbol{\phi})}
\bar\rho(\mathbf{x})\,\mathrm{d}\mathbf{x}\right)\phi_i \Bigg],
\end{multline}
with the $i$th Laguerre cell
\begin{multline}\label{eq:laguerre}
\Laguerre{i}(\mathbf{p},\boldsymbol{\phi}) := \Big\{\mathbf{x}\in\domain :
\tfrac{1}{2}\|\mathbf{p}_i-\mathbf{x}\|_2^2 - \phi_i \\
\le \tfrac{1}{2}\|\mathbf{p}_j-\mathbf{x}\|_2^2 - \phi_j, \ \forall j\neq i \Big\}.
\end{multline}
The mass and barycenter of each Laguerre cell,
\begin{equation}\label{eq:mass_bary}
a_i := \int_{\Laguerre{i}} \bar\rho(\mathbf{x})\,\mathrm{d}\mathbf{x},
\quad
\mathbf{b}_i := \frac{1}{a_i} \int_{\Laguerre{i}} \mathbf{x}\,\bar\rho(\mathbf{x})\,\mathrm{d}\mathbf{x},
\end{equation}
%play a central role in optimal transport-based coverage.
\reply{are stacked in $\mathbf{a}$ and $\mathbf{b}$, respectively, and} play a central role in optimal transport-based coverage.

\subsection{Coverage Control}
\label{sec:coverage}
Coverage control studies how to optimally position mobile agents over a domain weighted by a target density \cite{cortes2004coverage}. Formally, it seeks to design a dynamical system that drives the agent configuration $\agents$ toward a solution of $\min_{\agents} \mathcal{J}(\mathbf{p},\bar \rho)$, where $\mathcal{J}$ measures the quality of coverage with respect to the desired density $\densityd$.

While typically addressed via Voronoi tessellation~\cite{cortes2004coverage,lee2015multirobot}, Inoue et al.~\cite{inoue2020optimal} reformulated static coverage as the semi-discrete optimal transport problem
\begin{equation} \label{eq:objective_function_otcc}
\min_{\mathbf{p}(t)} \mathcal{W}_2^2(\mu_t, \bar \mu) = \min_{\mathbf{p}(t)} \max_{\boldsymbol \phi(t)} F(\mathbf{p},\boldsymbol{\phi}),
\end{equation}
where $\mu_t$ is the empirical measure of the agents at time $t$, $\bar \mu$ is the desired stationary measure, and $F$ is defined in~\eqref{eq:dual_objective}. The associated gradient descent-ascent dynamics
\begin{align}
\dot{\mathbf p}_i(t) &= -K_x \big(\mathbf{p}_i(t)-\mathbf{b}_i(t)\big), \label{eq:inoue_p}\\
\dot \phi_i(t) &= K_{\phi}\left(\frac{1}{N}- a_i(t)\right), \label{eq:inoue_phi}
\end{align}
drive the system toward the stationary saddle point $\mathbf p^\star = \mathbf b^\star$ with $\mathbf{a} = \mathbf{1}_N/N$, partitioning the domain into equal-mass regions and converging to their regions' barycenter at optimality. 
Unlike Voronoi partitions, which depend only on agent positions, Laguerre partitions incorporate the dual variables $\boldsymbol{\phi}$ to enforce equal-mass constraints and adapt directly to the target density, so that agents concentrate in high-density areas. When $\phi_i=\phi$ for all $i$, Laguerre regions reduce to Voronoi cells.
While Voronoi-based methods minimize geometric distortion without controlling the assigned mass, Laguerre partitions prioritize high-density regions and naturally arise from the minimization of the Wasserstein distance. Consequently, the OT framework yields a more accurate and structurally consistent approximation of the target distribution (see Sec.~\ref{sec:validation}).
The approach of~\cite{inoue2020optimal} is restricted to stationary target distributions. In this work, we extend the framework to time-varying target densities, so that the Laguerre partition becomes time-dependent and the agent configuration must track a moving saddle point.

\section{Modeling and Problem Statement}
\label{sec:problem}
We consider the time-varying coverage problem, in which a multi-agent system tracks a spatially distributed, time-varying target density by minimizing the instantaneous Wasserstein distance, on the basis of information about the agent states, the target density, and its time derivative~\cite{inoue2020optimal,cortes2004coverage,lee2015multirobot}.
The agents evolve in \reply{$\Rn^d$}, while the target density $\densityd : \domain \times \Rn_{\geq 0} \to \Rn_{> 0}$ has compact convex support $\domain$ and induces an absolutely continuous target measure $\bar\mu_t \in \mathcal{P}_2(\domain)$.

\begin{assumption}\label{th:assumption_ref}
$\densityd$ is $C^1$ and Lipschitz in time.
\end{assumption}

The $i$th agent is located at $\agent{i} \in \reply{\mathbb{R}^d}$, with $\agents := \mathrm{vec}(\agent{1}, \dots, \agent{N})$ collecting all positions. Each agent is fully actuated,
\begin{equation}\label{eq:agent_model}
\vagent{i} = \mathbf{u}_i(t), \quad i = 1, \dots, N,
\end{equation}
with $\mathbf{u}_i(t) \in \Rn^d$ the control input, and the empirical measure of the ensemble at time $t$ is $\mu_t := \tfrac{1}{N}\sum_{i=1}^N \delta_{\agent{i}}$.

\begin{assumption}\label{ass:assumption_agents}
Agents never overlap, i.e., $\agent{i} \neq \agent{j}$ for all $i \neq j$ and all $t \geq 0$.
\end{assumption}

\begin{assumption}\label{ass:sensing}
Each agent $i$ has a sensing and communication range sufficient to evaluate $\bar\rho(\mathbf{x},t)$ and $\partial_t \bar\rho(\mathbf{x},t)$ at any point of its Laguerre cell $\Laguerre{i}$ and to exchange state information with its Laguerre neighbors, \reply{i.e., $(\mathbf{p}_j , \phi_j)$ for $j \in \mathcal{N}_i(t)$, where $\mathcal{N}_i(t)$ is the set of neighbors of agent $i$ at time $t$. Consequently, agent $i$ can reconstruct its Laguerre cell $\Laguerre{i}$ and compute $a_i(t)$ and $\mathbf b_i(t)$.}
\end{assumption}

Assumption~\ref{ass:assumption_agents} is standard~\cite{cortes2005,inoue2020optimal} and can be enforced by adding a collision-avoidance term to~\eqref{eq:agent_model}. Although some works explicitly account for collisions~\cite{hussein2007effective,kim2024area}, this is beyond the scope of the present work. Assumption~\ref{ass:sensing} is also standard and 
the required information may be obtained through direct sensing, model-based estimation, or communication~\cite{inoue2020optimal,cortes2004coverage,lee2015multirobot,cortes2005spatially}.

The objective is to design $\mathbf{u}_i(t)$ for $i = 1, \dots, N$ so that, at each time $t$, the agent positions minimize the mismatch between the empirical and target distributions:
\begin{equation}\label{eq:problem}
\min_{\agents} \WH{2}.
\end{equation}

\begin{remark}\label{rem:benamou_brenier}
Problem~\eqref{eq:problem} performs instantaneous Wasserstein minimization, in contrast to the Benamou-Brenier formulation~\cite{benamou2000computational}, which optimizes over an entire time horizon in the mean-field setup.
\end{remark}

\section{Time-varying Optimal Transport Strategy}
\label{sec:solution}
Consider the Laguerre regions $\Laguerre{i}:=\Laguerre{i}(\agents, \boldsymbol{\phi}(t))$ for $i=1,\dots,N$ defined in \eqref{eq:laguerre}. Via the dual Kantorovich formulation, \eqref{eq:problem} can be recast as a min-max problem with explicit time dependence:
\begin{equation}
\min_{\agents} \max_{\boldsymbol{\phi}(t)} F(\agents, \boldsymbol{\phi}(t), t).
\end{equation}
This setting introduces challenges absent in the stationary case. First, purely feedback-based controllers \cite{inoue2020optimal} exhibit tracking lag when the optimum evolves. We address this by adding a feedforward term that compensates for the temporal variation of the target density, recovering exponential tracking guarantees consistent with time-varying optimization results \cite{davydov2025time}. 
Second, the evolution of $\bar \rho(x,t)$ \reply{indirectly} alters the Laguerre tessellation, with cell boundaries undergoing topological transitions at isolated times, potentially causing discontinuities in the closed-loop dynamics, requiring a Carath\'eodory solution framework.
\reply{Third, exact asymptotic tracking requires centralized computation. To overcome this limitation, we derive decentralized approximations that retain the structure of the centralized controller while relying only on local information.}

\subsection{Proposed solution}
To track a time-varying primal-dual stationary point of the instantaneous dual problem, we propose continuous-time dynamics for both the agent positions $\agents$ and the dual variables $\boldsymbol{\phi}(t)$.

Under Assumption~1, the objective in~\eqref{eq:problem} is continuously differentiable with respect to $(\mathbf{p}, \boldsymbol{\phi})$~\cite{cortes2005,bourne2015centroidal,merigot2011multiscale}, with partial derivatives
\begin{equation}\label{eq:partial_pi}
\frac{\partial F}{\partial \mathbf{p}_i} = a_i(t)\bigl(\agent{i} - \mathbf{b}_i(t)\bigr), \quad
\frac{\partial F}{\partial \phi_i} = \frac{1}{N} - a_i(t),
\end{equation}
obtained via the Reynolds transport theorem, since boundary terms cancel between adjacent Laguerre cells~\cite{bourne2015centroidal,inoue2020optimal}. Setting~\eqref{eq:partial_pi} to zero yields a critical point $(\mathbf{p}^\star(t), \boldsymbol{\phi}^\star(t))$ with
\begin{subequations}\label{eq:saddle_point}
\begin{align}
\mathbf{p}_i^\star(t) &= \mathbf{b}_i(t), \label{eq:min} \\
\phi_i^\star(t) &\in \Bigl\{\phi_i(t) \;\big|\; a_i(t) = \tfrac{1}{N} \Bigr\}. \label{eq:max}
\end{align}
\end{subequations}
Equation~\eqref{eq:saddle_point} shows that, at optimality, the agents induce an \emph{equal-mass} partition of the domain and track the \emph{barycenters} of their respective Laguerre regions. As the target density evolves, both the Laguerre cells and their barycenters change over time, giving rise to a time-varying coverage problem.

\reply{
For two adjacent Laguerre cells, define the common facet of dimension $d-1$ between agent $i$ and agent $j$, $\Gamma_{ij} := \bar{\Laguerre{i}} \cap \bar{\Laguerre{j}}$, and derive its normal outward velocity
\begin{align}
    \mathbf{v}_{ij} = \frac{(\mathbf{x}-\mathbf{p}_{i})^\top \dot{\mathbf{p}}_{i} + (\mathbf{p}_{j}-\mathbf{x})^\top \dot{\mathbf{p}}_{j} + \dot{\phi}_i(t) - \dot \phi_j(t)}{\lVert \mathbf{p}_{i} - \mathbf{p}_{j} \rVert}.
\end{align}
By the chain rule for derivatives and Reynolds theorem,
\begin{align*}
    \dot a_i(t) &= \partial_t a_i(t) + \sum_{j \in \mathcal{N}_i} \int_{\Gamma_{ij}} \densityd \mathbf{v}_{ij}(\mathbf x,t) \de S, \\
    \dot{\mathbf{b}}_i(t) &= \partial_t \mathbf{b}_i(t) \notag \\&+ \frac{1}{a_i(t)}\sum_{j \in \mathcal{N}_i} \int_{\Gamma_{ij}} (\mathbf{x}-\mathbf{b}_i(t)) \densityd \mathbf{v}_{ij}(\mathbf{x},t) \de S,
\end{align*}
where $\mathcal{N}_i(t)$ is the set of neighbors of agent $i$ in the Laguerre sense at time $t$, and where
\begin{subequations} \label{eq:FF_terms}
\begin{align}
    \partial_t a_i &= \int_{\Laguerre{i}} \partial_t \densityd \de \mathbf{x}, \\
    \partial_t \mathbf{b}_i &= \frac{1}{a_i(t)}\int_{\Laguerre{i}} (\mathbf{x}-\mathbf{b}_i(t)) \partial_t \densityd \de \mathbf{x}.
\end{align}
\end{subequations}
Noticing that $\mathbf{v}_{ij}$ is linear in $\dot{\mathbf{p}}(t)$ and $\dot{\boldsymbol{\phi}}(t)$, we rewrite the stack of the mass and barycenter derivatives as
\begin{align}
    \dot{\mathbf{a}} &= \partial_t \mathbf{a} + \mathbf{S}_{\phi a} \dot{\boldsymbol{\phi}} + \mathbf{S}_{pa} \dot{\mathbf{p}}, \label{eq:dota} \\
    \dot{\mathbf{b}} &= \partial_t \mathbf{b} + \mathbf{S}_{\phi b} \dot{\boldsymbol{\phi}} + \mathbf{S}_{pb} \dot{\mathbf{p}}, \label{eq:dotb}
\end{align}
where the sparse sensitivity matrices are $\mathbf{S}_{\phi a} \in \mathbb{R}^{N \times N}$, with nonzero scalar entries
\begin{subequations} \label{eq:sensitivity_matrices}
\begin{align}
    (\mathbf{S}_{\phi a})_{ii} &:= \sum_{j \in \mathcal{N}_i}\int_{\Gamma_{ij}} \frac{\densityd}{\lVert \agent{i}-\agent{j} \rVert} \de S, \\
    (\mathbf{S}_{\phi a})_{ij} &:= -\int_{\Gamma_{ij}} \frac{\densityd}{\lVert \agent{i}-\agent{j} \rVert} \de S, \quad \forall j \in \mathcal{N}_i,
\end{align}
$\mathbf{S}_{p a} \in \mathbb{R}^{N \times dN}$, with nonzero $1 \times dN$ block entries 
\begin{align}
    (\mathbf{S}_{p a})_{ii} &:= \sum_{j \in \mathcal{N}_i}\int_{\Gamma_{ij}} \frac{\densityd (\mathbf{x}-\agent{i})^\top}{\lVert \agent{i}-\agent{j} \rVert} \de S, \\
    (\mathbf{S}_{p a})_{ij} &:= -\int_{\Gamma_{ij}} \frac{\densityd (\mathbf{x} - \agent{j})^\top}{\lVert \agent{i}-\agent{j} \rVert} \de S, \quad \forall j \in \mathcal{N}_i,
\end{align}
$\mathbf{S}_{\phi b} \in \mathbb{R}^{dN \times N}$ with nonzero $dN \times 1$ block entries
\begin{align}
    (\mathbf{S}_{\phi b})_{ii} &:= \frac{1}{a_i} \sum_{j \in \mathcal{N}_i}\int_{\Gamma_{ij}} \frac{\densityd (\mathbf{x}-\mathbf b_{i})}{\lVert \agent{i}-\agent{j} \rVert} \de S, \\
    (\mathbf{S}_{\phi b})_{ij} &:= - \frac{1}{a_i} \int_{\Gamma_{ij}} \frac{\densityd (\mathbf{x}-\mathbf b_{i})}{\lVert \agent{i}-\agent{j} \rVert} \de S, \quad \forall j \in \mathcal{N}_i,
\end{align}
and $\mathbf{S}_{pb} \in \mathbb{R}^{dN \times dN}$, with nonzero $dN \times dN$ block entries
\begin{align}
    (\mathbf{S}_{p b})_{ii} &:= \frac{1}{a_i}\sum_{j \in \mathcal{N}_i}\int_{\Gamma_{ij}} \frac{\densityd (\mathbf{x}-\mathbf{b}_i) (\mathbf{x} - \agent{i})^\top}{\lVert \agent{i}-\agent{j} \rVert} \de S, \\
    (\mathbf{S}_{p b})_{ij} &:= - \frac{1}{a_i} \int_{\Gamma_{ij}} \frac{\densityd (\mathbf{x}-\mathbf{b}_i) (\mathbf{x} - \agent{j})^\top}{\lVert \agent{i}-\agent{j} \rVert} \de S.
\end{align}
\end{subequations}
Note that the dual variables $\boldsymbol \phi$ are defined up to an additive constant, since only differences between dual variables of adjacent cells matter in the construction of the Laguerre regions. Therefore, we need to remove a decision variable to resolve this ambiguity and, without loss of generality, we constrain one dual variable to zero, e.g., $\phi_N=0$. Because the outer minimization is nonconvex, we focus on tracking first-order optimality conditions, as standard in this problem \cite{inoue2020optimal,cortes2004coverage,lee2015multirobot}. Throughout, we define the position and mass errors $\mathbf{e}_p(t):=\mathbf{p}(t)-\mathbf{b}(t)$ and $\mathbf{e}_a(t)=\mathbf{a}(t)-\mathbf{1}_N/N$, respectively.
}

\reply{
First, we relax Assumption \ref{ass:sensing} to provide a centralized solution that perfectly tracks the first-order optimality condition \eqref{eq:saddle_point}. We then discuss distributed approximations in Sec. \ref{sec:distributed}.
\begin{theorem}[Asymptotic tracking]
\label{thm:main}
Suppose that Assumptions~\ref{th:assumption_ref}--\ref{ass:assumption_agents} hold and that topological changes in the Laguerre tessellation along the considered trajectory occur only at a locally finite set of time instants. Let $\mathbf{E}_N:= \begin{bmatrix}
    \mathbf{I}_{N-1} & \mathbf{0}_{N-1},
\end{bmatrix}^\top \in \mathbb{R}^{N \times (N-1)}$
impose $\boldsymbol\phi=\mathbf{E}_N\hat{\boldsymbol\phi}$ with $\phi_N=0$, and define 
\begin{equation}
\label{eq:M}
\mathbf M :=
\begin{bmatrix}
\mathbf{I}_{dN}-\mathbf{S}_{pb} & -\mathbf{S}_{\phi b}\mathbf{E}_N \\
\mathbf{E}_N^\top \mathbf{S}_{pa} & \mathbf{E}_N^\top \mathbf{S}_{\phi a}\mathbf{E}_N
\end{bmatrix},
\end{equation}
where $\mathbf{S}_{pb}$, $\mathbf{S}_{\phi b}$, $\mathbf{S}_{pa}$, and $\mathbf{S}_{\phi a}$ are defined in~\eqref{eq:sensitivity_matrices}.
Assume that the matrix $\mathbf{M}^{-1}$ is well-defined. 
Then, the control inputs in \eqref{eq:agent_model} defined by the continuous-time dynamics 
\begin{equation}
\label{eq:tracking_dynamics}
\begin{bmatrix}
\dot{\mathbf p}\\
\dot{\hat{\boldsymbol\phi}}
\end{bmatrix} = \mathbf M^{-1}
\begin{bmatrix}
-K_x\mathbf{e}_p+\partial_t\mathbf b \\
- \mathbf{E}_N^\top \left( K_\phi \mathbf{e}_a +\partial_t\mathbf a \right)
\end{bmatrix},
\end{equation}
with $K_x,K_\phi>0$ and feedforward terms $\partial_t \mathbf{a}, \partial_t \mathbf{b}$ defined in \eqref{eq:FF_terms}, achieve exponential tracking of the first-order optimality conditions \eqref{eq:saddle_point} of Problem~\eqref{eq:problem}.
In particular, the errors $\mathbf{e}_p$ and $ \mathbf{e}_a$ converge exponentially to zero with rates $K_x$ and $K_\phi$, respectively, almost everywhere.
\end{theorem}
\begin{proof}
The dynamics are understood in the Carath\'eodory sense, hence, all differential identities in the following hold almost everywhere, namely, outside the locally finite set of topology-changing time instants.
The first block row of~\eqref{eq:tracking_dynamics} gives
\begin{equation} \label{eq:first_block_proof}
(\mathbf{I}_{dN}-\mathbf{S}_{pb})\dot{\mathbf p} - \mathbf{S}_{\phi b}\mathbf{E}_N\dot{\hat{\boldsymbol\phi}} = -K_x(\mathbf p-\mathbf b)+\partial_t\mathbf b.
\end{equation}
Since
\begin{equation} \label{eq:dotb_proof}
\dot{\mathbf b} = \mathbf{S}_{pb}\dot{\mathbf p} + \mathbf{S}_{\phi b}\mathbf{E}_N\dot{\hat{\boldsymbol\phi}} + \partial_t\mathbf b,
\end{equation}
subtracting \eqref{eq:dotb_proof} from \eqref{eq:first_block_proof}, we obtain $\dot{\mathbf e}_p = -K_x \mathbf{e}_p$ a.e., leading to exponential convergence of the agents' positions to their barycenters with rate $K_x$.
The second block row gives
\begin{align}
\mathbf{E}_N^\top \left( \mathbf{S}_{pa}\dot{\mathbf p} +  \mathbf{S}_{\phi a}\mathbf{E}_N \dot{\hat{\boldsymbol\phi}} \right) &= -\mathbf{E}_N^\top \left(K_\phi \mathbf e_a + \partial_t\mathbf a \right).
\end{align}
Using
\begin{equation}
\dot{\mathbf a} = \mathbf{S}_{pa}\dot{\mathbf p} + \mathbf{S}_{\phi a}\mathbf{E}_N\dot{\hat{\boldsymbol\phi}} + \partial_t\mathbf a,
\end{equation}
we obtain
\begin{equation}
\mathbf{E}_N^\top \dot{\mathbf e}_a = -K_\phi \mathbf{E}_N^\top \mathbf e_a.
\end{equation}
Thus, the first $N-1$ components of $\mathbf e_a$ satisfy the desired exponential dynamics. Since
\begin{equation}
\mathbf1_N^\top\mathbf e_a=0,
\quad
\mathbf1_N^\top\dot{\mathbf e}_a=0,
\end{equation}
the $N$-th component satisfies the same equation. Hence $ \dot{\mathbf e}_a=-K_\phi\mathbf e_a$ almost everywhere. Moreover, this implies
\[
a_i(t)=\frac1N+ \left(a_i(0)-\frac1N\right)e^{-K_\phi t}>0,
\]
hence, every Laguerre cell retains positive mass and its barycenter remains well defined.
Because the position error decays to zero, \(\Omega\) is compact and each \(\mathbf b_i(t)\in\Omega\), the agent positions remain bounded . 
Since the agent positions are uniformly bounded and every Laguerre cell has positive mass, all differences of dual variables are uniformly bounded. Hence, under the adopted gauge, the reduced dual variable remains bounded and the closed-loop system does not diverge.
\end{proof}
}
\begin{remark}\label{rmk:robustness}
In some applications, $\partial_t \densityd$ is available, such as environmental monitoring or formation control governed by a known partial differential equation.
When $\partial_t \densityd$ is not directly available, estimation errors in $\partial_t \densityd$ act as bounded disturbances.
\end{remark}

The nonsingularity assumption of $\mathbf{M}$ in Theorem~\ref{thm:main} is typically made in the related literature on time-varying coverage control, where analogous invertibility conditions arise and are discussed, e.g., in~\cite{lee2015multirobot}. Further analysis is beyond the scope of this manuscript and left for future work.

\reply{Next, let us provide a stronger result for the one-dimensional case, where global optimization is recovered up to permutations of the agents, and more explicit structure is revealed. Without loss of generality, agents are labeled according to their initial order and by Assumption \ref{ass:assumption_agents}, this ordering is preserved.}
\reply{
\begin{theorem}[One-dimensional result]
\label{thm:one_dimensional_global}
Consider the setting of Theorem~\ref{thm:main} specialized to the one-dimensional domain $\Omega=[L,R]$.
For $i=1,\dots,N-1$, define the equal-mass quantiles $q_i$ by
\begin{equation}
\int_L^{q_i(t)}\bar\rho(x,t)\,dx=\frac{i}{N},
\end{equation}
with $q_0(t)=L$ and $q_N(t)=R$.
Then, at the optimum, $z_i^\star=q_i$ and the Laguerre cells are intervals
$\Laguerre{i}=[z_{i-1},z_i]$, where
\begin{equation}
    z_i = \frac{p_i+p_{i+1}}2 + \frac{\phi_i-\phi_{i+1}}{p_{i+1}-p_i}, \quad i=1,\ldots,N-1. \label{eq:1d_interfaces}
\end{equation}
The unique optimal agent positions in the ordered configuration are therefore
\begin{equation}
\label{eq:quantile_barycenters}
p_i^\star(t) = N \int_{q_{i-1}(t)}^{q_i(t)} x\bar \rho(x,t)\,dx.
\end{equation}
Let $\boldsymbol\phi^\star(t)$ be the unique dual variables under the constraint $\phi_N=0$, for which the Laguerre boundaries are $q_1(t),\dots,q_{N-1}(t)$. Then, the optimum $(\mathbf p^\star(t),\boldsymbol\phi^\star(t))$ is the unique solution of the min-max problem for $\phi_N=0$ and in the ordered configuration, i.e., up to permutations of the agent labels. Thus, every entry in \eqref{eq:tracking_dynamics} is explicitly computable from one-dimensional integrals and nearest-neighbor interface data and the optimality residuals converge exponentially to zero.
\end{theorem} }
\reply{
\begin{proof}
The OT problem is concave in $\boldsymbol{\phi}(t)$ \cite{peyre2019computational}, but, in general, non-convex in $\mathbf{p}(t)$.
For ordered agents in one dimension, however, the optimal transport is monotone \cite{peyre2019computational}. Hence the mass assigned to the $i$th agent is the quantile interval $[q_{i-1},q_i]$. Since $\bar \rho(x,t)>0, \, \forall (x,t)$, quantiles are uniquely defined. In particular, the semi-discrete transport cost at $\boldsymbol{\phi}=\boldsymbol{\phi}^\star$ is
\[
    \mathcal W_2^2(\bar \mu_t,\mu_t) = \frac12\sum_{i=1}^N \int_{q_{i-1}}^{q_i} |x-p_i|^2\bar \rho(x,t)\,dx.
\]
Since each interval has mass $1/N$, this expression is minimized at its barycenter,  \eqref{eq:quantile_barycenters}. Expanding the squares around $p_i^\star$ gives
\[
    \mathcal W_2^2(\bar \mu_t,\mu_t)-\mathcal W_2^2(\bar \mu_t,\mu_t^\star) = \frac{1}{2N}\sum_{i=1}^N|p_i-p_i^\star|^2.
\]
Thus $\mathbf p^\star$ is the unique minimizer among ordered configurations, up to permutations of the labels. Notice, moreover, that at the optimum,
\[
    \nabla_p\mathcal W_2^2=\frac1N(p-b),
    \quad
    \nabla_{pp}^2\mathcal W_2^2=\frac1N \mathbf{I}_N.
\]
Hence, $\nabla_{\mathbf p\mathbf p}^2 \mathcal W_2^2(\bar \mu_t,\mu_t) \succ 0$, and the optimum is also a nondegenerate strict minimizer. The boundary between two neighboring Laguerre cells satisfies
\[
    \frac12|z_i-p_i|^2-\phi_i = \frac12|z_i-p_{i+1}|^2-\phi_{i+1}.
\]
Solving this equality gives
\[
    z_i = \frac{p_i+p_{i+1}}2 + \frac{\phi_i-\phi_{i+1}}{p_{i+1}-p_i},
\]
which yields~\eqref{eq:1d_interfaces}. 
Summing \eqref{eq:max} over the first $i$ cells gives
\begin{align*}
    \int_L^{z_i^\star}
    \bar\rho(x,t)\, \de x = \sum_{j=1}^i \int_{z_{j-1}^\star}^{z_j^\star}
    \bar\rho(x,t)\,\de x = \sum_{j=1}^i\frac1N = \frac{i}{N}.
\end{align*}
By definition, the quantile $q_i$ is the unique point satisfying the above.
Therefore, each boundary depends only on two neighboring positions and weights. The masses, barycenters, and all their derivatives are consequently obtained from one-dimensional cell integrals and neighboring boundary data.
Substitution in \eqref{eq:1d_interfaces} determines every difference $\phi_i^\star-\phi_{i+1}^\star$. 
The additional imposed constraint $\phi_N=0$ then determines $\boldsymbol\phi^\star$ uniquely.  Therefore $(\mathbf p^\star,\boldsymbol\phi^\star)$ is the unique ordered optimal solution, which is tracked by \eqref{eq:tracking_dynamics} according to Theorem \ref{thm:main}.
\end{proof}
}

\subsection{Distributed implementation} \label{sec:distributed}
\reply{
The result established in Theorem \ref{thm:main} allows us to prescribe specific dynamics for the position and mass errors. However, its implementation requires the inversion of the matrix \(\mathbf{M}\), which may be computationally cumbersome, particularly in higher-dimensional settings.
We next present a simpler, albeit weaker, result. The next method enables each agent to track its Laguerre barycenter via a numerically approximated feedforward term, while ensuring that the mass-balancing error remains bounded, thereby providing an approximate solution to the original problem. As demonstrated in Sec. \ref{sec:validation}, the method still achieves better approximation performance than existing state-of-the-art approaches and admits a natural decentralized implementation.}

\begin{theorem}[Gradient-based approximate tracking]
\label{thm:main_old}
Suppose Assumptions~\ref{th:assumption_ref}--\ref{ass:sensing} hold. \reply{Moreover, assume that (i) $a_i(t)>0, \, \forall i\in [1,N], t \geq 0$, (ii) topological changes in the Laguerre tessellation along the considered trajectory occur only at a locally finite set of time instants, and that (iii) the following quantities satisfy $\lambda_\phi:=\liminf_{t \to \infty} \lambda_2(\mathbf{S}_{\phi a}(t))>0$ for $\lambda_2(A)$ being the second smallest eigenvalue of matrix $A$ and $\bar d := \limsup_{t \to \infty} \lVert \mathbf{S}_{pa} \mathbf{u} + \partial_t \mathbf{a} \rVert_2<+\infty$.} The control inputs $\mathbf u:=\mathrm{vec}(\mathbf{u}_i,\dots,\mathbf{u}_N)$ in \eqref{eq:agent_model} defined by the continuous-time dynamics 
    \begin{subequations}
    \label{eq:solution_gdff_ga}
    \begin{empheq}[left=\empheqlbrace]{align}
        \dot{\mathbf{p}}_i(t) &= -K_x \bigl(\agent{i} - \mathbf{b}_i(t)\bigr) + \dot{\mathbf{b}}_i(t), \label{eq:gd_ff} \\
    \dot{\phi}_i(t) &= K_{\phi} \left(\dfrac{1}{N} - a_i(t)\right),\label{eq:ga_phii}
    \end{empheq}
\end{subequations}
    for $i = 1, \dots, N$, where $\dot{\mathbf{b}}_i$ is the ideal implicit total time derivative of $\mathbf{b}_i$, and $K_x, K_\phi > 0$ are the primal and dual gains, respectively, achieve \reply{exponential convergence to the barycentric first-order condition and boundedness of the mass residual characterizing the time-varying first-order optimality conditions of Problem~\eqref{eq:problem}: 
    \begin{align*}
        \lim_{t \to \infty} \lVert \mathbf{e}_p(t) \rVert_2=0, \quad
        \limsup_{t \to \infty} \left\| \mathbf{e}_a(t) \right\|_2 \leq \frac{\bar d}{K_\phi \lambda_\phi}. 
    \end{align*}
    }
\end{theorem}
\begin{proof}
The primal problem is in general nonconvex~\cite{lee2015multirobot}, and global optimality is not claimed; following~\cite{lee2015multirobot,inoue2020optimal,cortes2004coverage,cortes2005}, the focus is on convergence to local critical points that evolve in time. By~\eqref{eq:min}, any such points correspond to agents located at the barycenters of their Laguerre cells~\cite{du2006convergence}.
For nondegenerate configurations, $\mathbf{b}_i(t)$ depends smoothly on the agents' positions~\cite{cortes2008,merigot2011multiscale,liu2009centroidal}. 
Topology changes of the Laguerre diagram occur only at a \reply{locally finite set of time instants}.
Hence, $\mathbf{b}(t)$ is absolutely continuous along trajectories, $\dot{\mathbf{b}}(t)$ exists almost everywhere, and solutions of~\eqref{eq:solution_gdff_ga} are understood in the Carath\'{e}odory sense~\cite{cortes2008,Filippov1988}. 
Consider the Lyapunov candidate $V = \tfrac{1}{2}\|\mathbf{e}_p\|_2^2.$
Since both $\mathbf{p}(t)$ and $\mathbf{b}(t)$ are absolutely continuous, so is $V(t)$, and the choice \eqref{eq:gd_ff}
yields, almost everywhere, $\dot V = -2 K_x V$.
Therefore, the tracking error
$\|\mathbf{e}_p\|_2$ vanishes exponentially with rate $K_x>0$.
The function $F(\mathbf{p}^\star(t), \boldsymbol{\phi})$ is concave in $\boldsymbol{\phi}$ near $\boldsymbol{\phi}^\star$~\cite{aurenhammer1998minkowski}, so gradient-ascent drives the masses $a_i(t)$ toward $1/N$, yielding the dual update \eqref{eq:ga_phii}, as in~\cite{inoue2020optimal}. 
\reply{This alone in general, however, does not track a possibly time-varying optimal $\boldsymbol{\phi}^\star(t)$ instantaneously. 
The imposed dynamics for the error $\mathbf e_a$ is $\dot{\mathbf{e}}_a=\dot{\mathbf a}$ for $\dot{\mathbf{a}}$ described in \eqref{eq:dota}.
Taking as Lyapunov function $V_a = \frac{1}{2} \lVert \mathbf{e}_a \rVert^2_2$, yields
\begin{align*}
    \dot V_a& = -K_\phi \mathbf{e}_a^\top \mathbf{S}_{\phi a} \mathbf{e}_a + \mathbf{e}_a^\top \left( \mathbf{S}_{pa}\dot{\mathbf{p}} + \partial_t \mathbf{a}\right) \\&\leq -K_{\phi} \lambda_2(\mathbf{S}_{\phi a}) \lVert \mathbf{e}_a \rVert_2^2 + \lVert \mathbf{e}_a \rVert_2 \lVert \mathbf{S}_{pa}\dot{\mathbf{p}} + \partial_t \mathbf{a} \rVert_2.
\end{align*}
Because the masses sum to $1$, the smallest eigenvalue of $\mathbf{S}_{\phi a}$ is zero and the relevant smallest eigenvalue is $\lambda_2(\mathbf{S}_{\phi a})>0$ by assumption.
By the comparison lemma, the thesis follows.
Moreover, following the same steps as in Theorem~\ref{thm:main}, the closed-loop system does not diverge under the stated assumptions.
}
\end{proof}

A key challenge in the proposed strategy is the derivation of $\dot{\mathbf{b}}_i(t)$ with local information only. 
The expression in \eqref{eq:dotb} \reply{is implicit} and may be cumbersome to find due to the time-varying integration domains \cite{pimenta2009simultaneous}.
Hence, it is typically approximated numerically \cite{lee2015multirobot}.
\reply{In practice, agent $i$ does not evaluate $\dot{\mathbf{b}}_i$ through \eqref{eq:dotb}; instead, obtains an explicit estimate from the time history of $\mathbf{b}_i$ using finite differences or a dynamic differentiator, as detailed in the following remark.}

\begin{remark}\label{rmk:numerical}
In decentralized numerical implementations, the feedforward term can be approximated using local information in a distributed fashion, for instance via $\dot{\mathbf{b}}_i \approx \partial_t \mathbf{b}_i =  a_i^{-1} \int_{\Laguerre{i}} (\mathbf{x}-\mathbf{b}_i)\, \partial_t \densityd \de\mathbf{x}$, finite differences, or other numerical differentiation schemes, such as the sliding mode differentiator used in Sec.~\ref{sec:validation}. 
\reply{Defining the estimated barycenter total derivative $\hat{\dot{\mathbf{b}}}=\dot{\mathbf{b}} + \boldsymbol{\eta}_b$ for the bounded estimation error $\lVert \boldsymbol{\eta}_b \rVert \leq \varepsilon_b$, one can directly adapt Theorem \ref{thm:main_old} to find 
\begin{align*}
    \limsup_{t \to \infty} \lVert \mathbf{e}_p \rVert \leq \frac{\varepsilon_b}{K_x}, \quad
     \limsup_{t \to \infty} \lVert \mathbf{e}_a \rVert \leq \frac{\bar d}{K_\phi \lambda_\phi}.
\end{align*}
} 
Topology changes occur only at isolated instants and are interpreted in the Carath\'eodory sense. \reply{In the simulations, the Laguerre diagram was recomputed at every evaluation of the dynamics, so the updated adjacency was automatically used after each change without requiring event handling.}
\end{remark}

Alternatively, one can implement a decentralized approximation of \eqref{eq:tracking_dynamics} based on Neumann series as in \cite{lee2015multirobot}. Inverting the sparse matrix $\mathbf{M}$ in Theorem \ref{thm:main} suggests a centralized implementation, as $\mathbf{M}^{-1}$ is, in general, not sparse.
However, a decentralized approximation can be obtained by exploiting the TVD-D$_k$ method, based on truncated Neumann series and $k$-hop communication, as established in \cite{lee2015multirobot} for Voronoi-based coverage.
\reply{We can rewrite $\mathbf{M}=\mathbf{I}_{N(d+1)-1}-\hat{\mathbf{M}}$ for
\begin{align}
    \hat{M} := \begin{bmatrix}
        \mathbf{S}_{pb} & \mathbf{S}_{\phi b} \mathbf{E}_N \\
        -\mathbf{E}_N^\top \mathbf{S}_{pa} & \mathbf{I}_{N-1}-\mathbf{E}_N^\top \mathbf{S}_{\phi a} \mathbf{E}_N
    \end{bmatrix},
\end{align}
}
and exploit the Neumann-series expansion \cite{horn2012matrix}: for a matrix $A$ with spectral radius smaller than $1$, one has $(I-A)^{-1} = \sum_{i=0}^\infty A^i$.
One multiplication by \(\hat M\) requires each agent to exchange its current intermediate iterate only with its one-hop Laguerre neighbors. After \(k\) communication rounds, the resulting control input depends only on data propagated through the \(k\)-hop Laguerre neighborhood. Assuming that the spectral radius of $\hat{M}$ is less than $1$, this provides a decentralized approximation analogous to the Voronoi-based TVD-D\(_k\) method in~\cite{lee2015multirobot}.
For a more intuitive interpretation, consider the one-dimensional case. Here, $\hat{M}$ is block tridiagonal because each agent has at most one left and one right neighbor. Consequently, the $i$th block row of $\hat{M}^j$ depends only on agents whose indices are within distance $j$ of agent $i$.

\section{Numerical validation} \label{sec:validation}
\begin{figure*}[tbp]
    \centering
    \vspace{0.3cm}
    \subfloat[]
    {
    \includegraphics{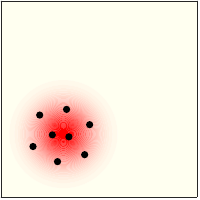}
    \label{fig:exp2D_formation_t1}
    }
    \subfloat[]    
    {\includegraphics{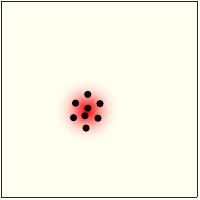}
    \label{fig:exp2D_formation_t2}
    }
    \subfloat[]    
    {\includegraphics{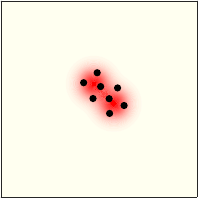}
    \label{fig:exp2D_formation_t4}
    }
    \subfloat[]    
    {\includegraphics{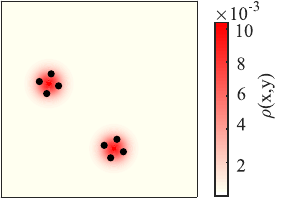}
    \label{fig:exp2D_formation_final}
    }
    \caption{Validation of the control law TVOT-G with $8$ agents (black dots) covering a time-varying density at (a) $t=1$, (b) $t=2$, (c) $t=4$, and (d) $t=6$ a.u. The agents maintain formation as the density shrinks, expands, and splits into a bimodal distribution.
    }
    \label{fig:formation_control}
\end{figure*}

\begin{figure}[htb]
    \centering
    \subfloat[]    
    {\includegraphics{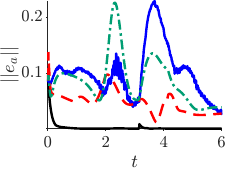}
    \label{fig:error_mass_2D}
    } 
    \hfill
    \subfloat[]    
    {\includegraphics{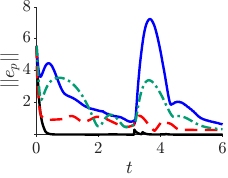}
    \label{fig:error_b_2D}
    }
    \caption{\reply{(a) Mass and (b) position errors over time for the 2D representative example using TVOT-C (black solid lines); TVOT-D$_4$ (dashed red lines); TVOT-D$_1$ (dotted green lines); and TVOT-G (solid blue lines).}}
    \label{fig:errors_2D}
\end{figure}

The proposed strategy is first validated in a 2D domain with a representative example comprising $8$ agents.
Fig.~\ref{fig:formation_control} shows four snapshots of the example using the controller described in Theorem \ref{thm:main_old}, which performs a gradient ascent on the dual variables and gradient descent with numerically estimated feedforward on the position (termed TVOT-G). 
Here, the feedforward term in \eqref{eq:solution_gdff_ga}, $\dot{\mathbf{b}}(t)$, is estimated online via a second-order sliding mode differentiator \cite{levant2003higher}:
\begin{equation}
    \begin{cases}
        \dot{\mathbf{z}}_0 = -\lambda_1 \lvert \mathbf{z}_0 - \mathbf{b} \rvert^{1/2} \mathrm{sign}(\mathbf{z}_0 - \mathbf{b}) + \mathbf{z}_1, \\
        \dot{\mathbf{z}}_1 = -\lambda_2 \, \mathrm{sign}(\mathbf{z}_0 - \mathbf{b}),
    \end{cases}
\end{equation}
yielding $\dot{\mathbf{b}}(t) \approx \mathbf{z}_1(t)$. 
The target density begins as a unimodal Gaussian with linearly varying mean and sinusoidally varying variance (Fig.\ref{fig:exp2D_formation_t1}), causing the agents to contract and expand accordingly, while tracking the moving mean (Fig.\ref{fig:exp2D_formation_t2}).
After one period, the distribution splits into a bimodal Gaussian (Fig.\ref{fig:exp2D_formation_t4}); the agents then divide into two groups to cover the emerging modes and continue tracking as the modes separate (Fig.\ref{fig:exp2D_formation_final}). Throughout the experiment, the agents maintain coherent spatial arrangements while adapting to the evolving density.
This experiment also demonstrates robustness to a violation of Assumption~\ref{th:assumption_ref} at an isolated time instant. Although the target density is nondifferentiable at the switching instant $t=3.14$~a.u., the proposed strategy continues to operate without significant degradation.
After a brief transient, the Wasserstein error decreases and reaches a plateau as the variances remain constant (omitted here for brevity, see supplementary materials).
\reply{The other controllers, i.e., the centralized controller described in Theorem \ref{thm:main} (TVOT-C) and its decentralized approximations based on Neumann series described in Sec. \ref{sec:distributed} (TVOT-D$_k$), reproduce qualitatively similar behavior, although reaching different critical points due to the non-convexity of the minimization problem, and have comparable Wasserstein errors over time (see supplementary materials).}
Notice that allowing the target density to vary over time also leads naturally to a formulation of formation control in probability space. 
Instead of prescribing relative positions among agents, the formation is implicitly encoded by the target distribution, whose regions of high density define the shape and evolution of the formation. By driving agents to effectively approximate this density over time, the collective behavior naturally maintains coherent spatial arrangements that adapt smoothly to changes in the desired formation, environment, or task requirements.
\reply{Fig. \ref{fig:errors_2D} shows the mass and position errors associated with the first-order optimality conditions and compares the performance of the proposed controller across several implementations. TVOT-G exhibits the largest peak errors, although its qualitative tracking performance remains satisfactory, as demonstrasted in Fig. \ref{fig:formation_control}. This baseline is compared with the centralized perfect-tracking solution TVOT-C and with two decentralized approximations based on truncated Neumann series using, respectively, $k=1$ and $k=4$.
As predicted by Theorem \ref{thm:main}, TVOT-C causes the mass and barycenter errors to converge exponentially to zero. Small transient spikes occur only near the time instant at which the reference density is nondifferentiable; nevertheless, both errors rapidly return to zero. TVOT-D$_1$ preserves the desired qualitative behavior but results in noticeably larger errors. Increasing the local information to $k=4$ hops reduces both errors, yielding performance much closer to that of the centralized solution.}

\begin{figure*}[htb]
    \centering
    \subfloat[]
    {\includegraphics[scale=0.99]{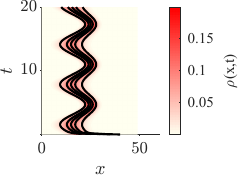}
    \label{fig:exp1D_traj}}
    \hfill
    \subfloat[]
    {\includegraphics[scale=0.99]{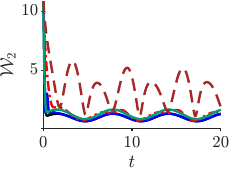}
    \label{fig:exp1D_W}}
    \hfill
    \subfloat[]
    {\includegraphics[scale=0.99]{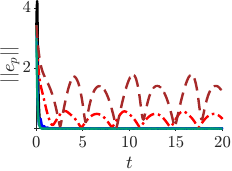}
    \label{fig:exp1D_e_b}}
    \hfill
    \subfloat[]
    {\includegraphics[scale=0.99]{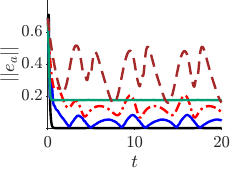}
    \label{fig:exp1D_e_a}}
    \caption{Validation in one dimension with $5$ agents. (a) Density evolution (background color from white to red) and agent trajectories using TVOT-C (solid black lines) for the sinusoidally varying mean and variance reference density. (b) Wasserstein error $\mathcal{W}_2$; \reply{(c) position error norm $\lVert e_b\rVert$; and (d) error mass norm $\lVert e_a\rVert$}, for TVOT-C (black solid lines); TVOT-D$_1$ (dashed red lines); TVOT-G (solid blue lines); OTCC \cite{inoue2020optimal} (solid green lines); TVV-C \cite{lee2015multirobot} (dashed brown lines).}
    \label{fig:exp1D}
\end{figure*}

We now validate a 1D case, where both the mean and variance vary sinusoidally: $\bar \rho(x,t) = \mathcal{N}(m(t), \sigma^2(t))$ with $\dot{m}(t) = v_m \sin(t)$ and $\dot{\sigma^2}(t) = v_{\sigma} \sin(t)$. The $N=5$ agents compress when the variance decreases and spread out when it increases, tracking the oscillating mean without lag (Fig.~\ref{fig:exp1D}). The Wasserstein distance oscillates in response to the varying variance (Fig.~\ref{fig:exp1D_W}, black line). 

The proposed strategies, i.e., \reply{centralized time-varying optimal transport (TVOT-C); the distributed counterpart via Neumann approximation with $k=1$ hop (TVOT-D$_1$); and the gradient-based solution with feedforward on position only described in Theorem \ref{thm:main_old}, where $\dot b_i$ is estimated via finite differences (TVOT-G)}, are compared and benchmarked against two baselines: quasi-static optimal transport coverage control (OTCC)~\cite{inoue2020optimal} and centralized time-varying Voronoi-based control (TVV-C)~\cite{lee2015multirobot}.
Without the feedforward term, the agents lag behind the moving density, resulting in significantly larger OTCC residual errors. All the proposed solutions also outperform the Voronoi-based approaches in Wasserstein error, highlighting the advantage of optimal-transport-based partitions in approximating continuous density distributions with a finite number of agents. Among the proposed methods, the centralized controller achieves the best overall performance, although the improvement over the decentralized solution and the gradient-based solution with numerically estimated feedforward is relatively small -- $2.34\%$ and $7.41\%$, respectively (Fig. \ref{fig:exp1D_W}).
\reply{
The Voronoi-based controller, having a feedforward term, is able to track its corresponding Voronoi barycenters -- but they do not track the Laguerre barycenters. The gradient-based controller, which includes a feedforward term on the agents' positions, instead accurately tracks the Laguerre barycenters, up to an approximation error due to the finite difference scheme (see Remark \ref{rmk:numerical}). In contrast, OTCC exhibits large oscillations. TVOT-D$_1$ achieves satisfactory tracking performance, comparable with the centralized one, and with barycenter errors $72.7\%$ smaller than those of the OTCC controller without feedforward (Fig. \ref{fig:exp1D_e_b}).
A similar behavior is observed for the mass error. The main difference is that, in this case, only the centralized solution achieves zero error, consistently with the theoretical results. TVOT-G exhibits bounded oscillations in the cell masses, while TVOT-D$_1$ maintains comparatively small errors and provides a good approximation of the centralized behavior. As expected from having $K_{\phi}=0$, TVV-C converges to a nonzero mass-error plateau.
Quantitative tables and additional simulations are found in the supplementary materials.
}

\begin{figure}[tbp]
    \centering
    \subfloat[]{%
        \includegraphics[width=0.41\columnwidth]{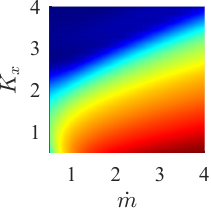}%
        \label{fig:comparison_ff}%
    }\hfill
    \subfloat[]{%
        \includegraphics[width=0.48\columnwidth]{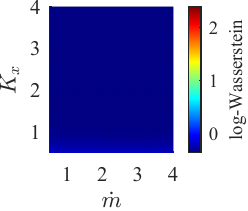}%
        \label{fig:comparison_tv}%
    }
    \caption{Wasserstein error (log scale) as a function of density drift velocity (x-axis) and gain $K_x$ (y-axis) for (a) OTCC without feedforward and (b) the proposed TVOT-G strategy.}
    \label{fig:comparison}
\end{figure}

The benefit of feedforward compensation is further quantified in Fig.~\ref{fig:comparison}, which compares TVOT-G and OTCC for varying density drift velocity $v_m \in [0.5, 4]$ and gain $K_x \in [0.5, 4]$ of a target Gaussian distribution. Without feedforward, acceptable accuracy requires large gains and slow drift; otherwise, persistent lag produces substantial steady-state errors. With the feedforward term, TVOT-G achieves significantly smaller errors across the entire parameter range. \reply{Here, after the transient, the disturbance in the mass dynamics vanishes and TVOT-G achieves asymptotically zero mass error, approaching the same moving stationary configuration as TVOT-C.}

\begin{figure}[tbp]
    \centering
    \subfloat[]{\includegraphics[width=0.49\columnwidth]{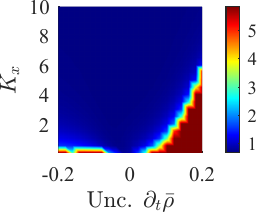}
    \label{fig:robustness}}
    \subfloat[]
    {\includegraphics[width=0.49\columnwidth]{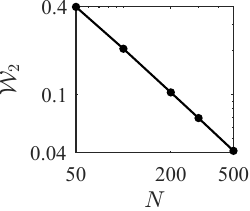}
    \label{fig:scaling}}
    \caption{(a) Robustness analysis showing Wasserstein error for varying estimation errors in $\partial_t \densityd$ (x-axis) and gain $K_x$ (y-axis); the target mean velocity is $5$ a.u. (b) Scaling analysis showing the $2$-Wasserstein error at final time versus number of agents $N$ (log-log scale); the slope of approximately $-0.99$ indicates $\mathcal{W}_2 \sim 1/N$.}
    \label{fig:robustness_scaling}
\end{figure}

Robustness to uncertainty in $\partial_t \bar \rho$ is evaluated in Fig.~\ref{fig:robustness}, where $\bar \rho=\mathcal{N}(m(t),\sigma^2)$, where perturbations of $\pm 20\%$ are introduced in the estimated velocity of the target density mean $\dot{m}=5$. Increasing $K_x$ reduces sensitivity to estimation errors, and tracking error remains bounded for a suitable range of parameters. Finally, scaling is examined by varying $N \in [50, 500]$. Fig.~\ref{fig:scaling} shows that the 2-Wasserstein error in 1D decays approximately as $1/N$, consistent with optimal quantization theory\cite{zador1982asymptotic}. 

\section{Conclusions}
\label{sec:conclusions}
This paper addressed coverage control for multi-agent systems tracking time-varying densities within a semi-discrete optimal transport framework. Coupled dynamics for the agent positions and the dual variables defining the Laguerre tessellation were derived, combining proportional feedback with feedforward compensation to track the time-varying barycenters while maintaining mass balance as the target density evolves.
Future work will focus on incorporating practical constraints such as collision avoidance, communication delays, and obstacles, on \reply{further comparing and characterizing the approximation error in} decentralized implementations based on Neumann truncation along the lines of~\cite{lee2015multirobot} \reply{with real-hardware experiments}, and on entropy-regularized formulations.

\begin{ack}
The authors wish to thank Davide Salzano, University of Naples Federico II, and Giovanni Russo, University of Salerno, for the useful discussions. AI tools were used solely to improve readability and revise language.
\end{ack}

\appendix

\section{Simulation parameters} 
All simulation parameters, quantitative comparison tables and additional simulations can be found at https://shorturl.at/vgPu8.

\bibliographystyle{plain}        % Include this if you use bibtex 
\bibliography{refs}           % and a bib file to produce the 

@book{villani2008optimal,
  title={Optimal transport: old and new},
  author={Villani, C{\'e}dric},
  volume={338},
  year={2008},
  publisher={Springer}
}

@book{peyre2019computational,
  title = {Computational Optimal Transport: {{With}} Applications to Data Science},
  author = {Peyr{\'e}, G. and Cuturi, M.},
  year = 2019,
  series = {Foundations and Trends in Machine Learning},
  publisher = {Now Publishers},
  isbn = {978-1-68083-550-2}
}

@article{cortes2008,
  author  = {Cort\'{e}s, Jorge},
  title   = {Discontinuous dynamical systems},
  journal = {IEEE Control Systems Magazine},
  volume  = {28},
  number  = {3},
  pages   = {36--73},
  year    = {2008}
}

@book{Filippov1988,
  author    = {Aleksei F. Filippov},
  title     = {Differential Equations with Discontinuous Righthand Sides},
  publisher = {Kluwer Academic Publishers},
  year      = {1988},
  series    = {Mathematics and Its Applications},
  volume    = {18}
}

@article{inoue2020optimal,
  title={Optimal transport-based coverage control for swarm robot systems: Generalization of the {Voronoi} tessellation-based method},
  author={Inoue, Daisuke and Ito, Yuji and Yoshida, Hiroaki},
  journal={IEEE Control Systems Letters},
  volume={5},
  number={4},
  pages={1483--1488},
  year={2021},
  publisher={IEEE}
}

@article{cortes2004coverage,
  title={Coverage control for mobile sensing networks},
  author={Cort\'es, Jorge and Mart{\'\i}nez, Sonia and Karatas, Timur and Bullo, Francesco},
  journal={IEEE Transactions on Robotics and Automation},
  volume={20},
  number={2},
  pages={243--255},
  year={2004},
  publisher={IEEE}
}

@article{lin2025heterogeneous,
  title={Heterogeneous Collaborative Pursuit via Coverage Control Driven by {Fokker}-{Planck} Equations},
  author={Lin, Ruoyu and Kim, Soobum and Egerstedt, Magnus},
  journal={IEEE Transactions on Robotics},
  year={2025},
  volume={41},
  number={},
  pages={3649-3668},
  publisher={IEEE}
}

@article{cortes2005spatially,
  title={Spatially-distributed coverage optimization and control with limited-range interactions},
  author={Cort\'es, Jorge and Mart{\'\i}nez, Sonia and Bullo, Francesco},
  journal={ESAIM: Control, Optimisation and Calculus of Variations},
  volume={11},
  number={4},
  pages={691--719},
  year={2005},
  publisher={EDP Sciences}
}

@article{lee2015multirobot,
  title={Multirobot control using time-varying density functions},
  author={Lee, Sung G. and Diaz-Mercado, Yancy and Egerstedt, Magnus},
  journal={IEEE Transactions on Robotics},
  volume={31},
  number={2},
  pages={489--493},
  year={2015},
  publisher={IEEE}
}

@article{chen2021optimal,
  title={Optimal transport in systems and control},
  author={Chen, Yongxin and Georgiou, Tryphon T and Pavon, Michele},
  journal={Annual Review of Control, Robotics, and Autonomous Systems},
  volume={4},
  number={1},
  pages={89--113},
  year={2021},
  publisher={Annual Reviews}
}

@article{du2006convergence,
  title={Convergence of the {Lloyd} algorithm for computing centroidal {Voronoi} tessellations},
  author={Du, Qiang and Emelianenko, Maria and Ju, Lili},
  journal={SIAM Journal on Numerical Analysis},
  volume={44},
  number={1},
  pages={102--119},
  year={2006},
  publisher={SIAM}
}

@book{horn2012matrix,
  title={Matrix Analysis},
  author={Horn, Roger A. and Johnson, Charles R.},
  year={2012},
  publisher={Cambridge University Press}
}

@article{bourne2015centroidal,
  title={Centroidal power diagrams, {Lloyd}'s algorithm, and applications to optimal location problems},
  author={Bourne, David P. and Roper, Steven M.},
  journal={SIAM Journal on Numerical Analysis},
  volume={53},
  number={6},
  pages={2545--2569},
  year={2015},
  publisher={SIAM}
}

@article{benamou2000computational,
  title={A computational fluid mechanics solution to the {Monge}-{Kantorovich} mass transfer problem},
  author={Benamou, Jean-David and Brenier, Yann},
  journal={Numerische Mathematik},
  volume={84},
  number={3},
  pages={375--393},
  year={2000},
  publisher={Springer-Verlag Berlin/Heidelberg}
}

@article{krishnan2024distributed,
  title={Distributed online optimization for multi-agent optimal transport},
  author={Krishnan, Vishaal and Mart{\'\i}nez, Sonia},
  journal={Automatica},
  volume={170},
  pages={111880},
  year={2024},
  publisher={Elsevier}
}

@inproceedings{bandyopadhyay2014probabilistic,
  title={Probabilistic swarm guidance using optimal transport},
  author={Bandyopadhyay, Saptarshi and Chung, Soon-Jo and Hadaegh, Fred Y},
  booktitle={IEEE Conference on Control Applications},
  pages={498--505},
  year={2014}
}

@article{liu2009centroidal,
  title={On centroidal {Voronoi} tessellation—energy smoothness and fast computation},
  author={Liu, Yang and Wang, Wenping and L{\'e}vy, Bruno and Sun, Feng and Yan, Dong-Ming and Lu, Lin and Yang, Chenglei},
  journal={ACM Transactions on Graphics},
  volume={28},
  number={4},
  pages={1--17},
  year={2009},
  publisher={ACM New York, NY, USA}
}

@inproceedings{pimenta2009simultaneous,
  title={Simultaneous coverage and tracking (SCAT) of moving targets with robot networks},
  author={Pimenta, Luciano C.A. and Schwager, Mac and Lindsey, Quentin and Kumar, Vijay and Rus, Daniela and Mesquita, Renato C. and Pereira, Guilherme A.S.},
  booktitle={Algorithmic Foundations of Robotics VIII},
  pages={85--99},
  year={2009},
  organization={Springer}
}

@article{krishnan2022multiscale,
  title={A multiscale analysis of multi-agent coverage control algorithms},
  author={Krishnan, Vishaal and Mart{\'\i}nez, Sonia},
  journal={Automatica},
  volume={145},
  pages={110516},
  year={2022},
  publisher={Elsevier}
}

@article{merigot2011multiscale,
  title={A multiscale approach to optimal transport},
  author={M{\'e}rigot, Quentin},
  journal={Computer Graphics Forum},
  volume={30},
  number={5},
  pages={1583--1592},
  year={2011},
  publisher={Wiley Online Library}
}

@article{kim2024area,
  title={Area coverage using multiple aerial robots with coverage redundancy and collision avoidance},
  author={Kim, Soobum and Lin, Ruoyu and Coogan, Samuel and Egerstedt, Magnus},
  journal={IEEE Control Systems Letters},
  volume={8},
  pages={610--615},
  year={2024},
  publisher={IEEE}
}

@article{hussein2007effective,
  title={Effective coverage control for mobile sensor networks with guaranteed collision avoidance},
  author={Hussein, Islam I. and Stipanovic, Dusan M.},
  journal={IEEE Transactions on Control Systems Technology},
  volume={15},
  number={4},
  pages={642--657},
  year={2007},
  publisher={IEEE}
}

@inproceedings{cortes2005,
author = {J. Cort\'es and S. Mart{\'\i}nez and T. Karatas and F. Bullo },
booktitle = {10th Mediterranean Conference on Control and Automation},
title = {Coverage control for mobile sensing networks: variations on a theme},
year = {2002},
address = {Lisbon, Portugal}
}

@article{aurenhammer1987power,
  title={Power diagrams: properties, algorithms and applications},
  author={Aurenhammer, Franz},
  journal={SIAM Journal on Computing},
  volume={16},
  number={1},
  pages={78--96},
  year={1987},
  publisher={SIAM}
}

@article{levant2003higher,
  title={Higher-order sliding modes, differentiation and output-feedback control},
  author={Levant, Arie},
  journal={International Journal of Control},
  volume={76},
  number={9-10},
  pages={924--941},
  year={2003},
  publisher={Taylor \& Francis}
}

@article{aurenhammer1998minkowski,
  title={Minkowski-type theorems and least-squares clustering},
  author={Aurenhammer, Franz and Hoffmann, Friedrich and Aronov, Boris},
  journal={Algorithmica},
  volume={20},
  number={1},
  pages={61--76},
  year={1998},
  publisher={Springer}
}

@article{davydov2025time,
  title={Time-varying convex optimization: A contraction and equilibrium tracking approach},
  author={Davydov, Alexander and Centorrino, Veronica and Gokhale, Anand and Russo, Giovanni and Bullo, Francesco},
  journal={IEEE Transactions on Automatic Control},
  volume={70},
  number={11},
  pages={7446-7460},
  year={2025},
  publisher={IEEE}
}

@article{zador1982asymptotic,
  title={Asymptotic quantization error of continuous signals and the quantization dimension},
  author={Zador, Paul},
  journal={IEEE Transactions on Information Theory},
  volume={28},
  number={2},
  pages={139--149},
  year={1982},
  publisher={IEEE}
}
                                 % bibliography (preferred). The
                                 % correct style is generated by
                                 % Elsevier at the time of printing.

\end{document}